\documentclass[acmtog, finalversion]{acmart}

\usepackage{booktabs} 

\usepackage[ruled]{algorithm2e} 

\SetAlFnt{\small}
\SetAlCapFnt{\small}
\SetAlCapNameFnt{\small}
\SetAlCapHSkip{0pt}
\IncMargin{-\parindent}

\acmJournal{TOG}
\acmYear{2017}\acmVolume{36}\acmNumber{4}\acmArticle{126}\acmMonth{7}
\acmDOI{10.1145/3072959.3073629}

\begin{document}
\title{DeepSketch2Face: A Deep Learning Based Sketching System for\\ 3D Face and Caricature Modeling}
\author{Xiaoguang Han}
\author{Chang Gao}
\author{Yizhou Yu}
\affiliation{%
  \institution{The University of Hong Kong}
  \department{Department of Computer Science}
  \city{Hong Kong}
}

\renewcommand\shortauthors{Han, X. et al}

\begin{teaserfigure}
 \centering
   \includegraphics[height=1.5in]{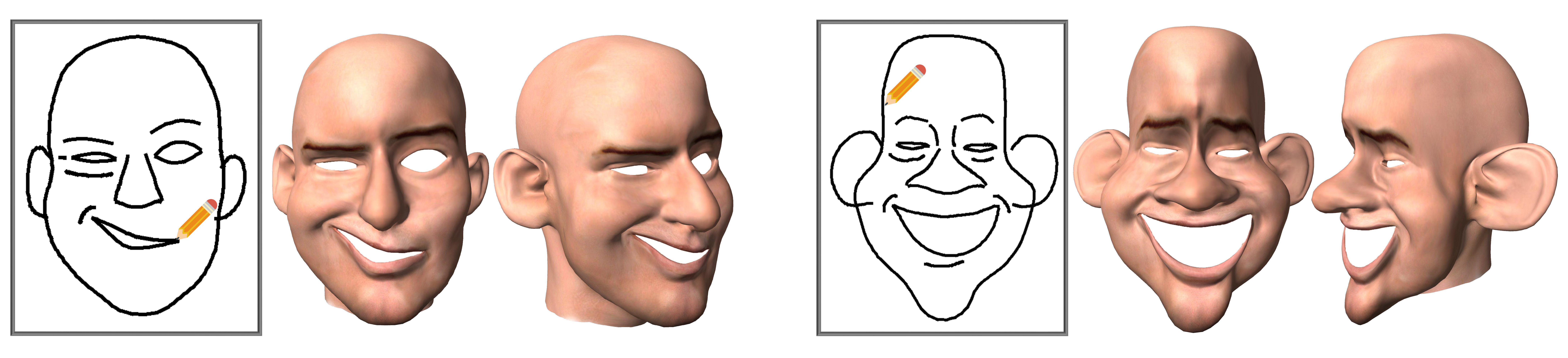}
   \caption{Using our sketching system, an amateur user can create 3D face or caricature models with complicated shape and expression in a few minutes. Both models shown here were created in less than 10 minutes by a user without any prior drawing and modeling experiences.}
   \label{fig:teaser}
\end{teaserfigure}

\begin{abstract}
Face modeling has been paid much attention in the field of visual computing. There exist many scenarios, including cartoon characters, avatars for social media, 3D face caricatures as well as face-related art and design, where low-cost interactive face modeling is a popular approach especially among amateur users. In this paper, we propose a deep learning based sketching system for 3D face and caricature modeling. This system has a labor-efficient sketching interface, that allows the user to draw freehand imprecise yet expressive 2D lines representing the contours of facial features. A novel CNN based deep regression network is designed for inferring 3D face models from 2D sketches. Our network fuses both CNN and shape based features of the input sketch, and has two independent branches of fully connected layers generating independent subsets of coefficients for a bilinear face representation. Our system also supports gesture based interactions for users to further manipulate initial face models. Both user studies and numerical results indicate that our sketching system can help users create face models quickly and effectively. A significantly expanded face database with diverse identities, expressions and levels of exaggeration is constructed to promote further research and evaluation of face modeling techniques.
\end{abstract}

%
%


%
%


\maketitle

\section{Introduction}
Face modeling has been a popular research topic in the field of visual computing, and different approaches, including real face digitization and interactive face modeling, have been attempted in response to a variety of application scenarios. While high-end applications, such as virtual characters in feature films, demand high-fidelity face models acquired from the real world, there exist many scenarios, including cartoon characters, custom-made avatars for games and social media, 3D face caricatures as well as face-related art and design, where low-cost interactive modeling is still a mainstream approach especially among amateur users.
Nevertheless, faces have rich geometric variations due to diverse identities and expressions, and interactively creating a decent 3D face model is a labor-intensive and time-consuming task even for a skilled artist with the help of well-developed software such as MAYA or ZBrush. Thus creating expressive face models with a minimal amount of labor is still a major challenge in interactive face modeling.

Let us leave 3D faces for a moment and think about how we draw a 2D face. It is a natural choice for us to quickly draw the outermost silhouette first and then the contours of various facial features including eyes, nose and mouth. Such a set of silhouettes and contours already give a very good depiction of the underlying face even though they are merely a sparse set of lines. Note that such silhouettes and contours also exist on a 3D face except that they have extra depth information and the silhouette is view-dependent. However, even when the 3D silhouette and contours are fully specified, the shape of the facial regions in-between these sparse lines are still unknown. Fortunately, the 3D shape of these regions might be highly correlated with the 2D and 3D shape of the silhouette and contours. For example, there exist strong correlations between the nose contour and the overall shape of the nose, and also between the mouth contour and the shape of the cheeks. Therefore, it is crucial to learn such correlations from data and let them guide labor-efficient face modeling.


Inspired by the above observations, in this paper, we propose a deep learning based sketching system for 3D face and caricature modeling for amateur users. In our system, a user only needs to draw or modify a sparse set of 2D silhouettes and facial feature contours. Sketching allows the user to draw freehand imprecise yet expressive 2D lines, giving rise to pleasant user experiences. Given the 2D facial sketch, an underlying deep learning based regression network automatically computes a corresponding 3D face model. Our system supports two phases of sketching, initial sketching and follow-up sketching. During initial sketching, the user draws a 2D sketch from scratch and the generated initial 3D model may not exactly match the sketch. At the beginning of follow-up sketching, feature contours on the initial 3D model are projected onto the 2D canvas to produce an updated sketch, and the user can redraw some of the projected lines if they are not satisfactory. The redrawn lines in this sketching phase become harder constraints that projected contours should match more closely.

In our system, an underlying deep regression network is responsible for converting a 2D face sketch to a 3D face model. Although a face has a fixed number of main features (such as eyes, nose and mouth), to maintain a flexible sketching interface, it is important to allow the user to draw an unspecified number of lines. To this end, we treat the 2D sketch as an image, and rely on a convolutional neural network (CNN) running on the raw pixels to compute a fixed-length description of all lines in the sketch. However, even a neuron in the topmost convolutional layer has a limited receptive field and cannot `see' the entire sketch. To compensate for the lack of global context, we also directly represent the overall shape of main facial features using a dimension reduction model. The output from our deep network is a set of coefficients for a bilinear face representation, which is able to reconstruct all 3D vertices of a face mesh. This bilinear representation considers identities and expressions as two independent modes with separate coefficients. To avoid interference between these two modes, our deep network has two independent branches with different numbers of fully connected layers following the shared convolutional layers.

To support research and evaluation of face modeling techniques including our deep learning based regression network, we prepare a large database containing both 3D face models and their associated 2D sketches. Our database is a significantly expanded version of the face database reported in \cite{cao2014facewarehouse}. To model 3D face caricatures, we use each original face model from \cite{cao2014facewarehouse} to create three new face models with different levels of exaggeration. 14 new facial expressions are also designed and transferred to all created models while similar facial expressions in the original database were merged. The resulting new database has 15,000 registered 3D face models, which include 150 identities, 25 expressions and 4 levels of exaggeration. Handdrawn 2D sketches of a subset of 3D face models are also included.

In summary, this paper has the following contributions.
\begin{itemize}
  \item A novel sketching system is proposed for 3D face and caricature modeling. This system has a labor-efficient sketching interface, and initial 3D face models can be automatically generated from 2D sketches through learned correlations between them.
      Our system supports gesture based interactions for users to further manipulate initial face models.
  \item A novel CNN based deep regression network is designed for inferring 3D face models from 2D sketches. Our network fuses both CNN and shape based features of the input sketch, and also has two independent branches of fully connected layers generating two independent subsets of coefficients for a bilinear face representation.
  \item A significantly expanded face database with diverse identities, expressions and levels of exaggeration is also constructed for training and testing. This database will be publicly released to benefit other researchers working on face modeling.
\end{itemize}

\begin{figure*}[ht]
  \includegraphics[width=0.98\textwidth]{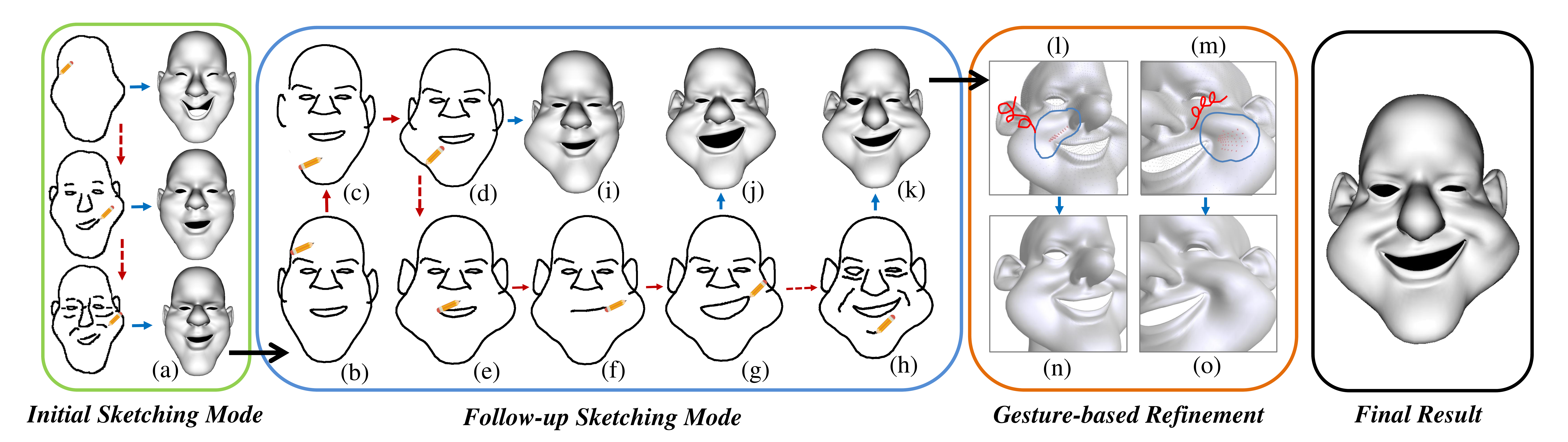}
  \caption{Our sketching system has three interaction modes: the initial sketching mode, follow-up sketching model and gesture-based refinement mode. In the initial sketching mode, the 3D face is updated immediately after each operation. The follow-up sketching mode gets started when an output model (a) in the initial sketching model is rendered to a sketch (b). A sequence of operations in this model are shown from (b) to (h). Users can switch in real time from 2D sketching to 3D model viewing (e.g. (d) to (i), (g) to (j) and (h) to (k)). The created shape (k) can be refined in the gesture-based refinement mode. (l) and (m) show the gestures used for depth depressing and bulging, and the corresponding results after each operation are shown in (n) and (o). A red solid arrow indicates a single operation while a dashed one means several operations, and a blue arrow stands for model updating.}\label{fig:workflow}
\end{figure*}

\section{Related Work}
\paragraph{Data-Driven Modeling by Sketching} When there is a database of 3D models, a sketch can be used as a query to search for the most similar shapes~\cite{eitz2012sketch,li20163d}.
Fan {\em et al.}~\shortcite{fan2013modeling} developed a system where retrieved models are shown along with the sketch as shadows to provide guidance to amateur users in creative modeling. Sketches can also help retrieve part candidates for assembling a man-made object~\cite{xie2013sketch}) or an entire indoor scene~\cite{xu2013sketch2scene}. While traditional machine learning algorithms often have a hard time robustly handling rough and noisy sketches created by amateur users, deep convolutional neural networks (CNNs)~\cite{krizhevsky2012imagenet} have been applied successfully to sketch understanding in image retrieval~\cite{sangkloy2016sketchy} and 3D shape retrieval~\cite{wang2015sketch}.

Recently, CNNs have also been used to learn a regression model mapping sketches to parameters for procedural modeling of man-made objects~\cite{huang2016shape} or buildings~\cite{Nishida:2016:ISU:2897824.2925951}. Incorporating CNN based parameter inference, Nishida {\em et al.}~\shortcite{Nishida:2016:ISU:2897824.2925951} also developed a user interface for real-time urban modeling. 
Major differences exist between our work and the work in \cite{huang2016shape} other than they deal with different types of objects. First, this paper presents a labor-efficient user interface that incorporates real-time sketching for interactive modeling and deep learning based gesture recognition for shape refinement while the modeling process adopted in ~\cite{huang2016shape} is not interactive and their proposed algorithm simply generates a 3D model from a complete drawing. Second, our network architecture has crucial novel components with respect to the architecture used in \cite{huang2016shape}. For example, our network has two independent branches of fully connected layers, and it also has a vertex loss layer which directly calculates vertex position errors. Both components effectively reduce model prediction errors. Third, this paper publicly releases a dataset while \cite{huang2016shape} did not. The dataset used in this paper has better connections with the real world because part of the 3D face models were originally digitized from real people and part of the sketches are real drawings collected from artists while the dataset used in \cite{huang2016shape} was entirely generated from 3D virtual models.

\paragraph{Morphable Face Models} Blanz and Vetter~\shortcite{blanz1999morphable} applied a PCA based representation learned from a dataset of 200 face models to single-view 3D face reconstruction. Multi-linear tensor decomposition is performed in \cite{vlasic2005face} to decouple variations of expression, identity and viseme into separate modes and encode a face model with subsets of independent coefficients. Tena {\em et al.}~\shortcite{Tena:2011} proposed a linear piecewise face modeling technique based on PCA sub-models that are independently trained but share boundaries. Recently, Cao {\em et al.}~\shortcite{cao2014facewarehouse} built a large face database with 150 identities and 20 expressions using an RGBD sensor and further trained a bilinear face representation. These morphable face representations have been widely used in applications such as 3D face reconstruction for image or video editing~\cite{yang2011expression,dale2011video} and real-time performance capture from RGB input~\cite{cao20133d,cao2016real,saito2016real} or RGBD input~\cite{weise2011realtime,bouaziz2013online}. In addition, there exist interactive face modeling systems~\cite{gunnarsson2007statistically,feng2008real,lau2009face}, which consider both the intention of users and statistical rules learned from an underlying database.
In our proposed system, other than the overall shape information provided by sketched lines, we also extract deep CNN features from 2D sketch images. Such deep features significantly improve the accuracy of morphable model inference. To train our deep regression network, we significantly expanded the database reported in \cite{cao2014facewarehouse} to include multiple levels of exaggeration in both identity and expression.

\paragraph{3D Face Caricatures} A face caricature shows facial features in an exaggerated manner. Creating 3D face models in this style is also a popular topic in computer graphics. A common approach exaggerates an input 3D model by magnifying the differences between the input model and a reference model (~\cite{lewiner2011interactive,vieira2013three}) through deformation. Sela {\em et al.}~\shortcite{sela2015computational} considered gradients as a measure of characteristics and proposed a method to perform exaggeration by scaling the gradients at mesh vertices.
Liu {\em et al.}~\shortcite{liu2009semi} proposed a machine learning method to map 2D feature points detected in natural images to the coefficients of a PCA model learned from a dataset of 200 3D caricature models. An interactive technique is proposed in \cite{xie2009interactive}, where each mouse operation on vertices triggers an inference of coefficients for PCA subspaces representing individual face components. Given an input face photo and its caricature drawing, the method proposed in \cite{clarke2011automatic} captures the deformation style of the caricature and apply the style to other photos. To our knowledge, the system in this paper is the first one that creates 3D face caricatures from 2D sketches.

\paragraph{Sketch-Based Freeform Modeling} Creating freeform 3D models from handdrawings has been paid much attention for two decades in both the industry and research community~\cite{olsen2009sketch}. Humans typically rely on silhouettes or contours to depict objects in drawings. Singh and Fiume~\cite{singh1998wires} proposed to use curves to control 3D object deformation. Igarashi {\em et al.}~\shortcite{Igarashi:1999:TSI:311535.311602} presented a system for automatically generating a 3D shape from 2D silhouettes interactively drawn by the user. Fibermesh~\cite{nealen2007fibermesh} adds another feature that allows the user to modify the geometry by moving selected curve handles on the surface. A sketched line near a silhouette can be used to define a deformation for shape refinement~\cite{nealen2007sketch}. Rivers {\em et al.}~\shortcite{rivers20103d} proposed a system to create 3D shapes from multi-view silhouettes.  The sketches in all these methods are used to constrain vertex positions on a 3D model. Recent work~\cite{shao2012crossshade,xu2014true2form,pan2015flow}) also studies the relationship between surface normals and 2D cross lines, and automatically generates surfaces from concept sketches. In our system, sketched lines not only serve as position constraints for silhouettes and contours but also help determine the 3D coordinates of other vertices according to complex correlations learned by our deep regression network.

\section{User Interface}
We introduce our system from the perspective of users in this section. The input device can be either a mouse or a pen tablet, and a pen tablet is recommended. Our system supports the following three interaction modes for coarse-to-fine face modeling: initial sketching mode, follow-up sketching mode and gesture-based refinement mode. A complete interactive modeling session in our system only requires drawing, erasing and clicking.

\begin{figure}[h]
  \includegraphics[width=0.95\columnwidth]{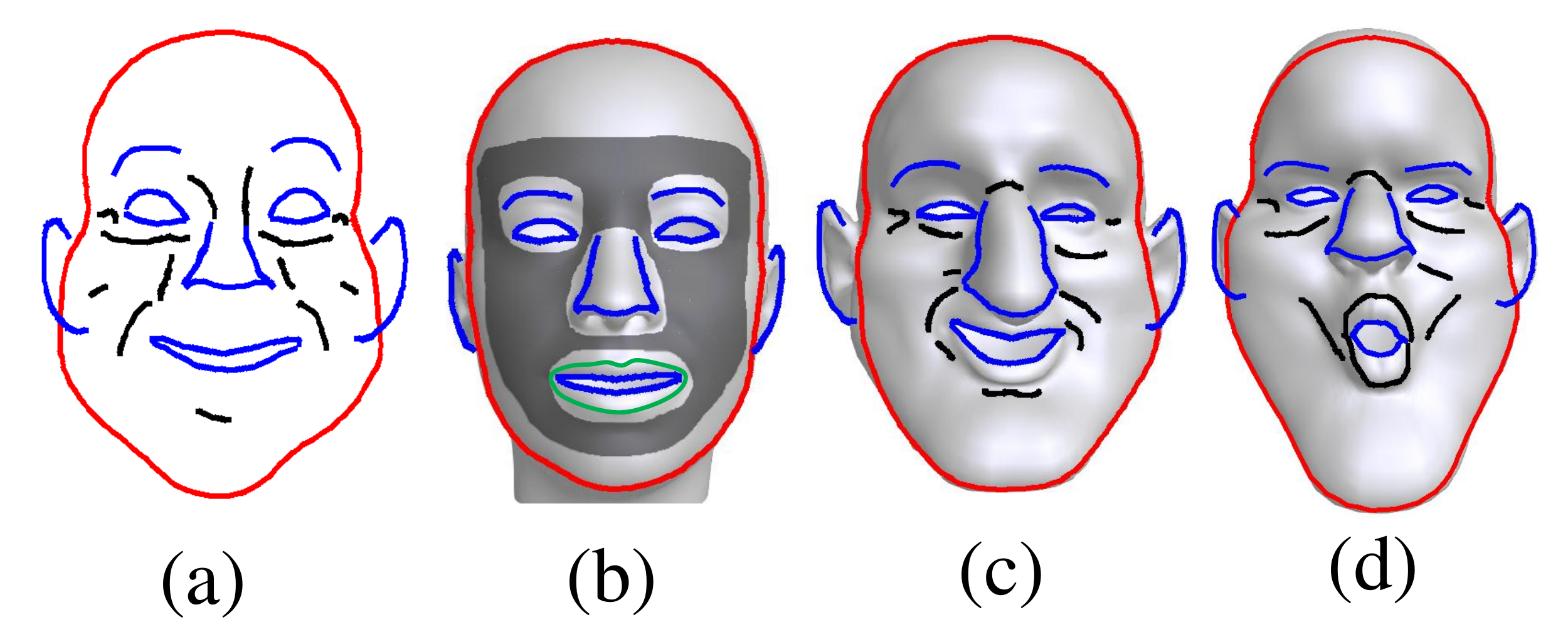}
  \caption{(a) The lines in a face sketch are separated into \emph{the silhouette line}(red), \emph{feature lines}(blue) and \emph{wrinkle lines} (black). (b) The silhouette line and feature lines are manually mapped to sequences of vertices on a template mesh, which can be used as curve handles during deformation. Suggestive contours in the region marked by dark gray are treated as wrinkle lines. As suggested by artists, the outer contour of the lips (green) is also treated as a wrinkle line (e.g. the one shown in (d)). (c) and (d) show the lines on two models with different shape and expression.}\label{fig:lines}
\end{figure}

We separate lines in a face sketch into three categories, \emph{silhouette line}, \emph{feature lines} and \emph{wrinkle lines}. As illustrated in Figure~\ref{fig:lines} (a), a silhouette line (highlighted in red) refers to the outermost contour of a face, \emph{feature lines} (highlighted in blue) include left and right eyebrows, the upper and lower contours of left and right eyes, the upper and lower contours of mouth, the silhouettes of left and right ears and the nose contour. All other lines are classified as \emph{wrinkle lines} (highlighted in black).

\subsection{Initial Sketching Mode}
In the initial sketching mode, a user can create a freehand face drawing on a blank canvas. The user can apply an arbitrary sequence of drawing and erasing operations as in a standard sketching interface. After every drawing or erasing operation, an updated 3D face corresponding to the latest sketch (no matter whether it is a complete face sketch or not) is automatically generated and displayed.
The initial sketching mode promotes completely unconstrained hand-drawing, which might be preferred by certain users. In this mode, a user completely new to our system can immediately start drawing without spending any time to know the components of our interface.
Note that the 2D projection of the feature contours on the 3D face generated in this mode is similar to the freehand drawing in both shape and expression, but they do not precisely match. The follow-up sketching mode in our system integrates deformation with model inference to achieve a better matching between the two.

\subsection{Follow-up Sketching Mode}
An important component of our user interface is a front-view sketching mode starting from a drawing consisting of 2D silhouette and feature lines projected from a 3D face model generated in the initial sketching mode. To revise this initial face model, the user can refine the drawing by editing or adding lines. Within this sketching mode, the user can switch in real time between 2D sketching and 3D model viewing. The face shown during 3D model viewing is generated from the latest 2D sketch on the fly. Note that a user can skip the initial sketching mode and directly start from the follow-up sketching mode by using a template face as the initial model.

The types of user interactions supported in this sketching model include {\em drawing}, {\em erasing} and {\em auto-snapping}. We describe our design of user interactions below and illustrate the process in Figure~\ref{fig:workflow} from (b) to (h).

Each feature or wrinkle line is represented as several neighboring but disconnected curve segments, and each segment is represented as a sequence of points. Each erasing operation removes one single segment or has no effect if there are no segments left. However, there are no predefined segments on a silhouette line, which is represented as a single closed sequence of points. An arbitrary part of this sequence can be erased as long as the erased part is connected. If multiple disconnected parts have been erased, our system locates intermediate points connecting the parts and remove them as well.

After an erasing operation has been performed on any silhouette or feature line, a drawing operation is required to replace the erased segment with a new one (the reason for this design is given in Section~\ref{sec:facemodeling}). An erased segment exposes a gap with two open endpoints on a silhouette or feature line, the latest drawn segment will be automatically snapped to fill the gap. For all other cases, the drawn strokes are saved as wrinkle lines.

As the lower and upper contours of mouth or eyes are always connected at the two corners, when one of the two contours is redrawn, a similar transformation is applied to the other to automatically snap it to the redrawn one.


\subsection{Gesture-Based 3D Face Refinement}
\begin{figure}[h]
  \includegraphics[width=0.95\columnwidth]{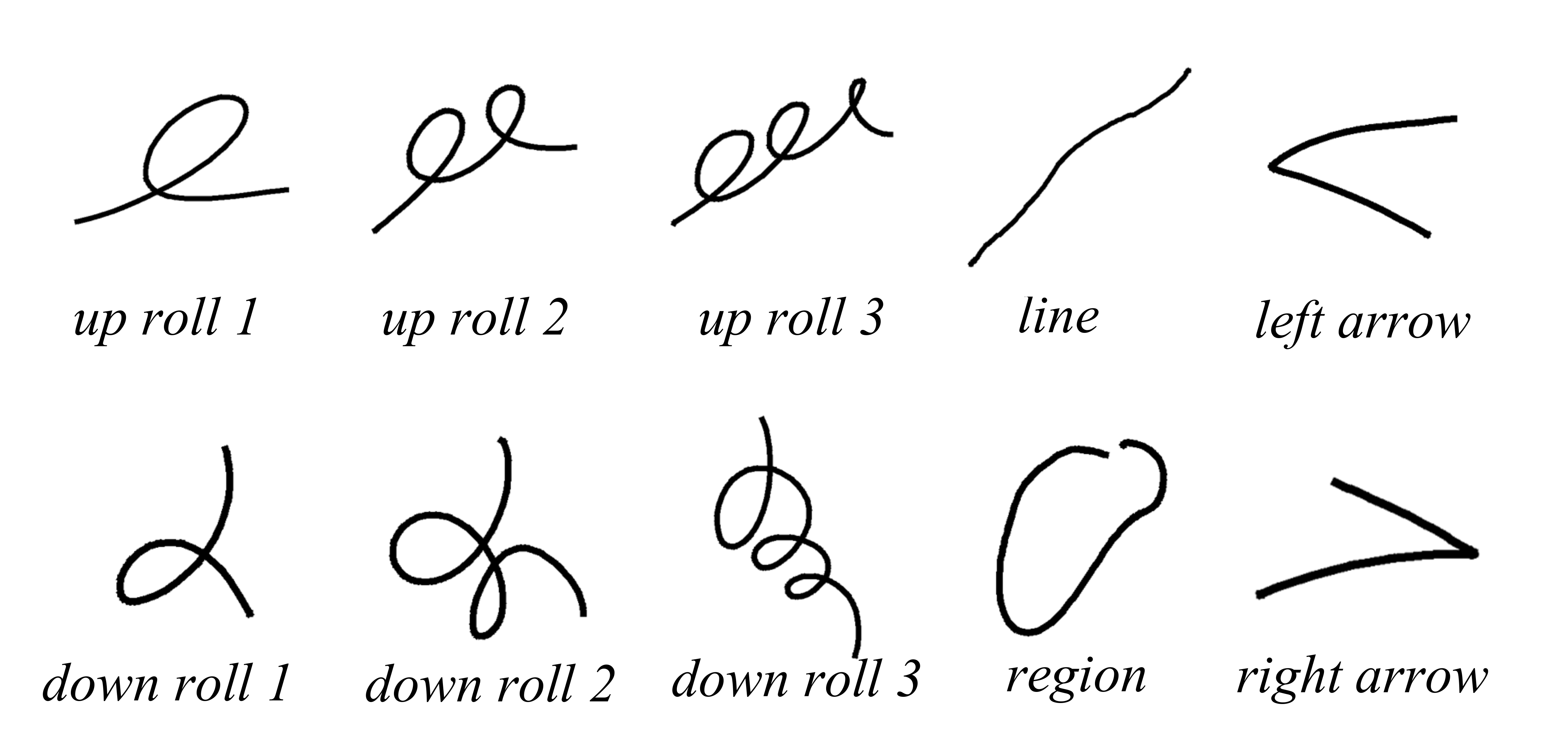}
  \caption{Gesture types in our refinement mode.}\label{fig:gesture}
\end{figure}

Once a 3D face model has been generated using 2D sketching, shape refinement can be further performed directly on the 3D model. Our shape refinement is built upon traditional handle-based deformation~\cite{sorkine2004laplacian}. We provide the same tool as in ~\cite{nealen2007sketch} for manipulating pre-defined silhouettes in left and right side view of a face model. Meanwhile, to make interactive refinement in an arbitrary view more intuitive and convenient, we define pen gestures and map them to specific operations in the model refinement mode. This is inspired by ~\cite{bae2008ilovesketch}. As shown in Figure~\ref{fig:gesture}, we design 10 different pen gestures named \emph{line}, \emph{region}, \emph{left arrow}, \emph{right arrow}, and three levels of \emph{up roll} and \emph{down roll}. \emph{left arrow} and \emph{right arrow} are mapped to undo and redo respectively, and the meanings of other gestures will be elaborated below.

Handle-based deformation has a selection step and a deformation step. In the selection step, the user can mark handle vertices by brushing and erasing. The deformation mode supports ROI selection, deformation in the XY-plane, deformation along the Z-axis and silhouette-driven deformation. First of all, a region of interest (ROI) needs to be defined prior to any deformation. Vertices outside this region are fixed. The \emph{region} gesture shown in Figure~\ref{fig:gesture} is naturally used to define the ROI. If ROI selection has never been performed, the entire mesh is set as the ROI.

Deformation in the XY-plane is triggered by a \emph{line} gesture in the deformation step. This gesture defines a 2D translation within the canvas (XY-plane), which is applied to the handle vertices. Translated handle vertices are set as position constraints in the deformation. When there are no handles marked, a \emph{line} gesture is interpreted as traditional silhouette-based shape editing~\cite{nealen2007sketch}. In this case, a silhouette is first identified on the mesh and a deformation is performed by dragging the silhouette towards the drawn line.

When a user wishes to modify the depth of handles with respect to the current view plane, traditionally, he or she needs to perform 2D deformation within another view plane perpendicular to the current one. However, in face modeling, the handle vertices are often occluded in such a perpendicular view. Our system relies on gestures to manipulate the depth of handle vertices directly. We map 3 levels of \emph{up roll} to 3 levels of bulging and 3 levels of \emph{down roll} to 3 levels of depression.

\begin{figure}[t]
  \includegraphics[width=0.95\columnwidth]{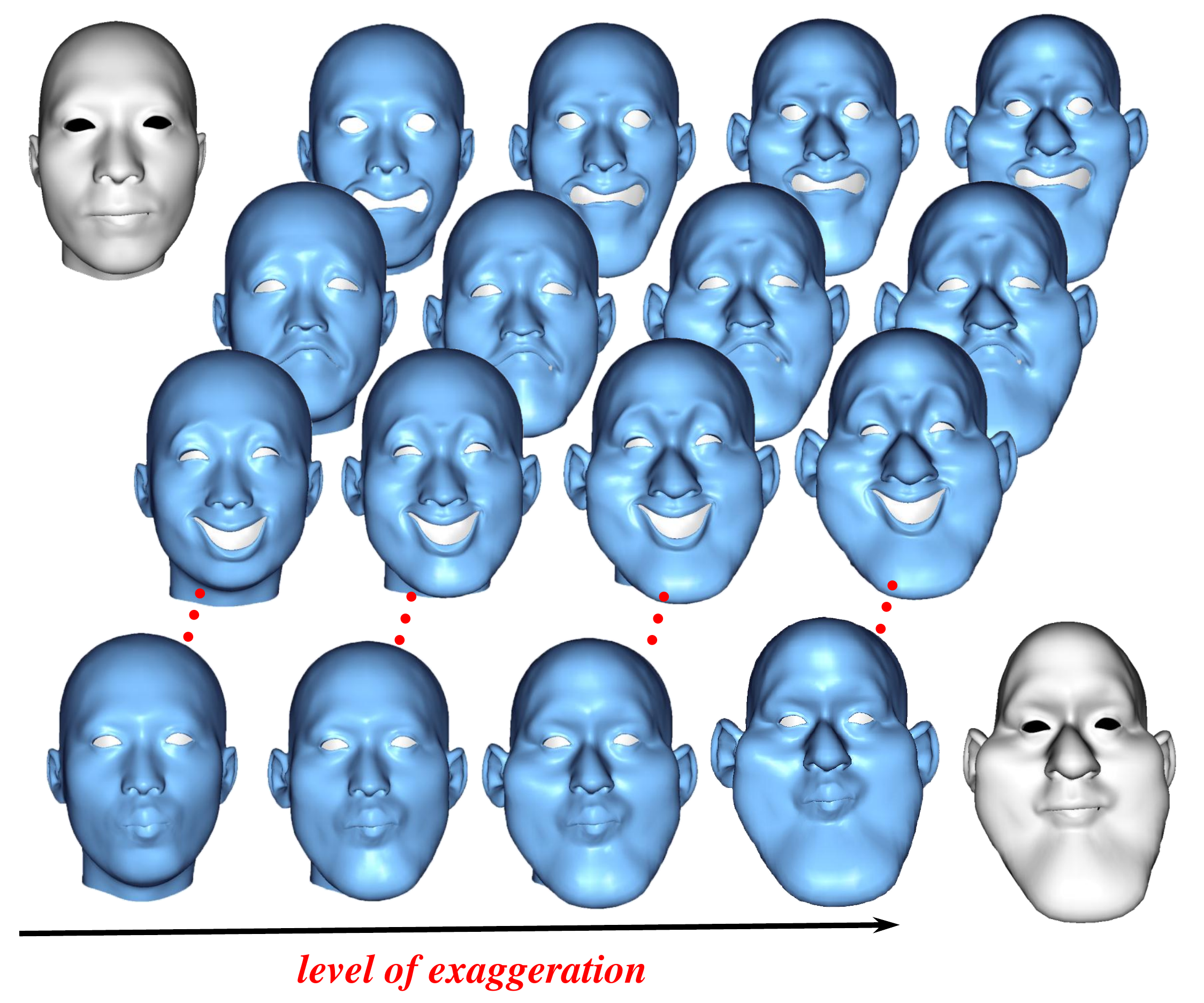}
  \caption{For the neutral face in the upper left corner, a subset of models with different expressions and levels of exaggeration in our dataset are shown. The corresponding neutral face with the highest level of exaggeration is shown in the lower right corner.}\label{fig:modelsample}
\end{figure}

\section{3D Face Inference from Sketches}
In this section, we present our method for performing 3D model inference from sketches. In essence, we use a deep neural network to approximate the function $V=\varphi(I)$ that maps a binary sketch image $I$ to the vertices of its corresponding 3D face model.

\subsection{Database Construction}
Let us first introduce our 3D face database and its construction process. Our database has 15,000 meshes, which include 150 identities, 25 expressions and 4 levels of exaggeration. It is a significantly expanded version of the public face database reported in \cite{cao2014facewarehouse}. Each mesh in the database has 11500 vertices. All meshes are well registered with exactly the same topology.

\paragraph{New Expressions} The database reported in \cite{cao2014facewarehouse} provides 20 facial expressions for each individual identity. However, many expressions only have minor differences from each other, which make their 2D sketches indistinguishable. To increase the diversity of expressions, we chose 11 expressions with relatively noticeable mutual differences from the original 20 expressions. In addition, we asked an artist to create 14 new expressions in caricature styles on a neutral face. These new expressions are associated with different moods, including ``happy", ``sad", and ``fearful". A subset of these 25 expressions are shown in Figure~\ref{fig:modelsample} and all of them can be found in the supplemental materials. The 14 new expressions are further transferred to all other identities using the deformation transfer technique from~\cite{sumner2004deformation}.

\paragraph{Shape Exaggeration} Shape exaggeration is based on the method in ~\cite{sela2015computational}. Given a mesh, it first applies a scaling factor to the gradients at vertices, and then reconstructs the 3D shape from these new gradients using the method in \cite{yu2004mesh}. Due to unstable gradients around eyes and mouth, we fix those few places and exaggerate other parts including cheeks, chin, forehead, nose and ears. A shape with level 1 exaggeration is actually the original shape. We apply three different scaling factors to shapes at level 1 to obtain shapes at higher levels of exaggeration. Sample results are shown in Figure~\ref{fig:modelsample}.

\begin{figure}[t]
  \includegraphics[width=0.95\columnwidth]{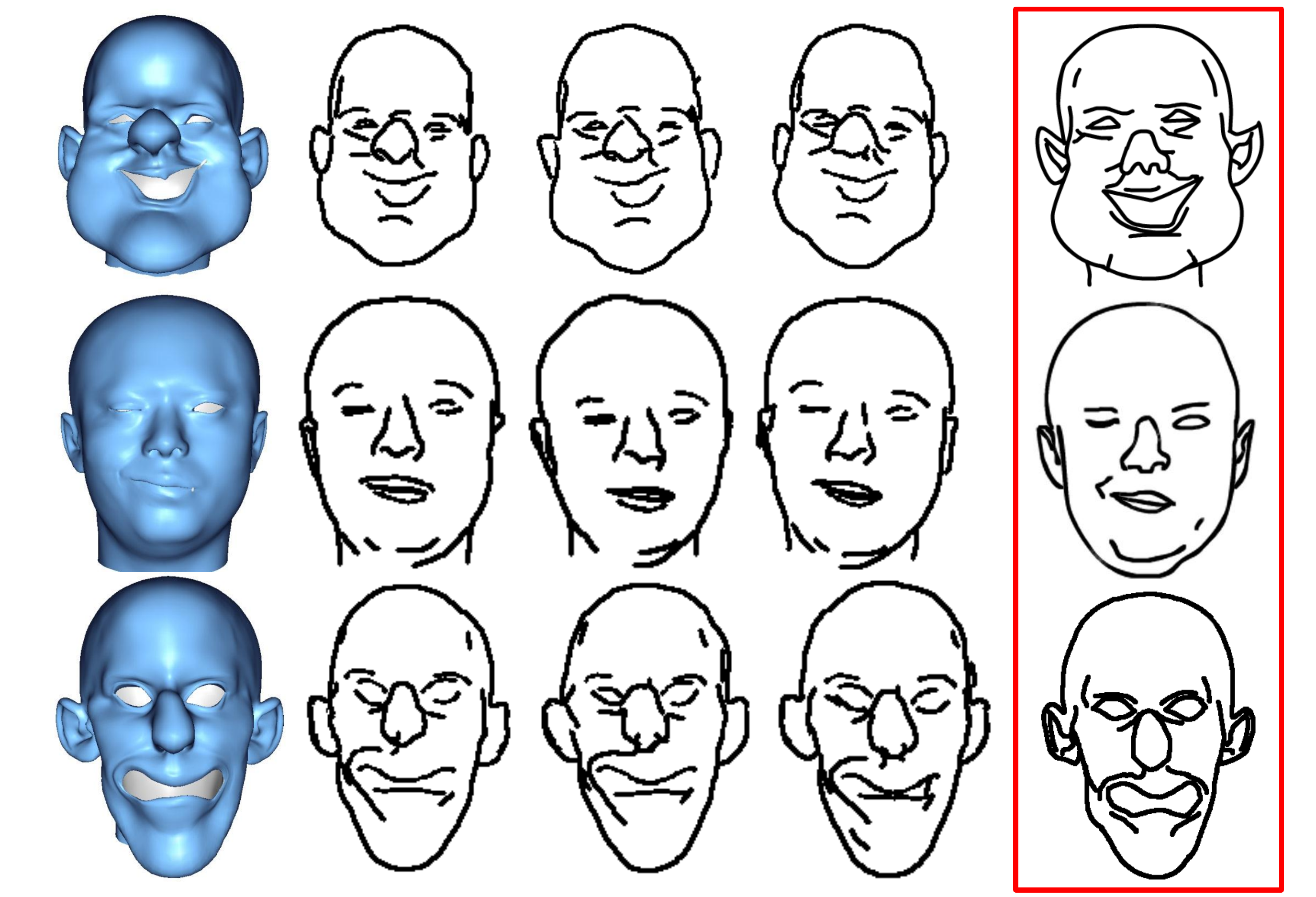}
  \caption{Each row shows three rendered sketches generated with our data augmentation schemes and a real sketch drawn by a user (highlighted in the red box).}\label{fig:sketchsample}
\end{figure}

\paragraph{Sketch Rendering} Another important part of our database is the set of 2D sketches corresponding to all 3D faces. On a template mesh, we predefine 3D curves representing the face silhouette and contours of facial features including eyes, nose and mouth. Each curve on the template mesh is a vertex sequence. A rendered sketch of each 3D face consists of a silhouette line, feature lines and wrinkle lines. We render the silhouette and feature lines directly from the predefined 3D curves, and render wrinkle lines using suggestive contours~\cite{decarlo2003suggestive}. Sample results are shown in Figure~\ref{fig:sketchsample}.

\paragraph{Hand-drawn Sketches} To make the sketches in our database closer to real hand drawings, we also collected real sketches drawn by users. 20 users with diverse drawing skills contributed to our hand-drawn data. Each user was given 200 rendered images of 100 3D faces, two images per face. These two images are respectively the frontal view and side view of the corresponding 3D face. The user was asked to draw a sketch to represent the shape and expression of each given 3D face. All 2000 3D faces used in this stage were randomly chosen from 15,000 models.

\begin{figure*}[t]
  \includegraphics[width=0.98\textwidth]{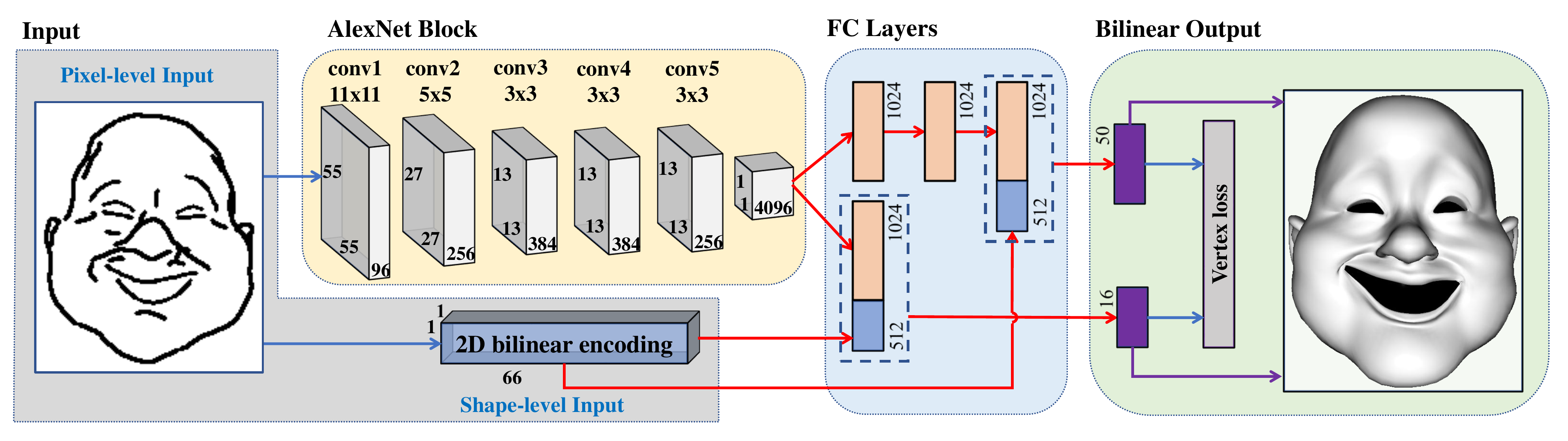}
  \caption{Our network architecture.}\label{fig:network}
\end{figure*}

\subsection{Bilinear Morphable Representation}
For the task of 3D face inference from sketches, we learn a bilinear morphable model for parametrically representing the subspace of shape and expression variations defined by our new database, as in \cite{cao2014facewarehouse}. We first construct a 3-mode tensor $T$ with 11,500 vertices $\times$ 600 identities $\times$ 25 expressions. Here, faces at different levels of exaggeration are viewed as different identities because an exaggerated face indeed has a different shape in comparison to the original face it was generated from and such differences in shape are independent of expression changes. After N-mode singular value decomposition and tensor truncation, a reduced core tensor $C$ is obtained. Then, any face mesh in our database with a vertex set $V$ can be approximated using an identity weight vector $u$ and an expression weight vector $v$ as follows.
\begin{equation}\label{eq:bilinear}
    V = C\times_{2}u^{T}\times_{3}v^{T},
\end{equation}
where the dimensions of $u$ and $v$ are set to 50 and 16 respectively in our experiments. In this representation, 3D face inference can be viewed as a regression function that maps a sketch image to the $u$ and $v$ vectors of its corresponding 3D face.

\subsection{Network Architecture}
The architecture of our deep regression network for 3D face inference is illustrated in Figure~\ref{fig:network}. It includes multiple convolutional layers to extract features from an input sketch image. These convolutional layers are set up in the same way as in AlexNet~\cite{krizhevsky2012imagenet}.

\paragraph{Bilinear Output} The output from our deep regression network is a set of coefficients for the bilinear face representation, which eventually reconstructs all 3D vertices of a face mesh. As discussed earlier, this bilinear representation considers identities and expressions as two independent modes with separate weight vectors $u$ and $v$. To avoid interference between these two modes, our deep network has two independent branches with different numbers of fully connected layers for generating the $u$ and $v$ vectors. However, we still keep the fully connected layer immediately following the convolutional layers in AlexNet.
The branch for generating the $u$ vector has three new fully connected layers while the branch for the $v$ vector only has one new fully connected layer as the complexity of the identity mode (600 shapes) is much higher than the complexity of the expression mode (25 types). Each fully connected layer in both branches has 1024 neurons, which was set empirically to achieve the optimal performance.

\paragraph{Pixel-Level Input} The first input to our deep regression network is a 256x256 binary sketch image with all silhouette, feature and wrinkle lines. This sketch image is fed to the convolutional layers in our network.

\paragraph{Shape-Level Input} Since even a neuron in the topmost convolutional layer has a limited receptive field and cannot `see' the entire sketch, the convolutional layers in our deep network might not be able to grasp global contextual information in the input sketch. To make full use of the information from the sketch, we directly sample a fixed number of points on the silhouette and feature lines and represent the 2D position of these sample points via another bilinear model, where the dimensions for the identity and expression modes are still set to 50 and 16. These 66 coefficients form our shape-level input vector. In each of the two branches, this shape-level input vector is followed by a new fully connected layer with 512 neurons, whose output is concatenated with the previous feature vector before the output layer. We have tried another alternative to concatenate the shape-level input vector directly with the feature vector from the remaining fully connected layer in the AlexNet block before they are fed into the two branches. This alternative scheme negatively affected the final performance.


\paragraph{Vertex Loss Layer} Our deep regression network is directly trained to minimize vertex approximation errors. For this purpose, we set up a vertex loss layer estimating the $L_2$ error between groundtruth vertices and vertices reconstructed from the predicted $u$ and $v$ vectors. This $L_2$ vertex approximation error is formulated as follows:
\begin{equation}\label{eq:vloss}
    E = \frac{1}{n}\sum_{i}w_i||C_i\times_{2}u^{T}\times_{3}v^{T}-g_i||^{2},
\end{equation}
where $C_i$ is the 2D slice of the core tensor $C$ corresponding to the $i$-th vertex, $g_i$ stands for the position of the $i$-th vertex on the groundtruth mesh, $w_i$ means the weight for the $i$-th vertex, and $n$ is the total number of vertices. This layer directly follows the layer predicting the $u$ and $v$ vectors. $w_i$'s are set according to the relative importance of vertices as certain vertices such as those on 3D silhouette and feature lines have higher importance.

\subsection{Network Training}

\begin{figure}[h]
  \includegraphics[width=0.95\columnwidth]{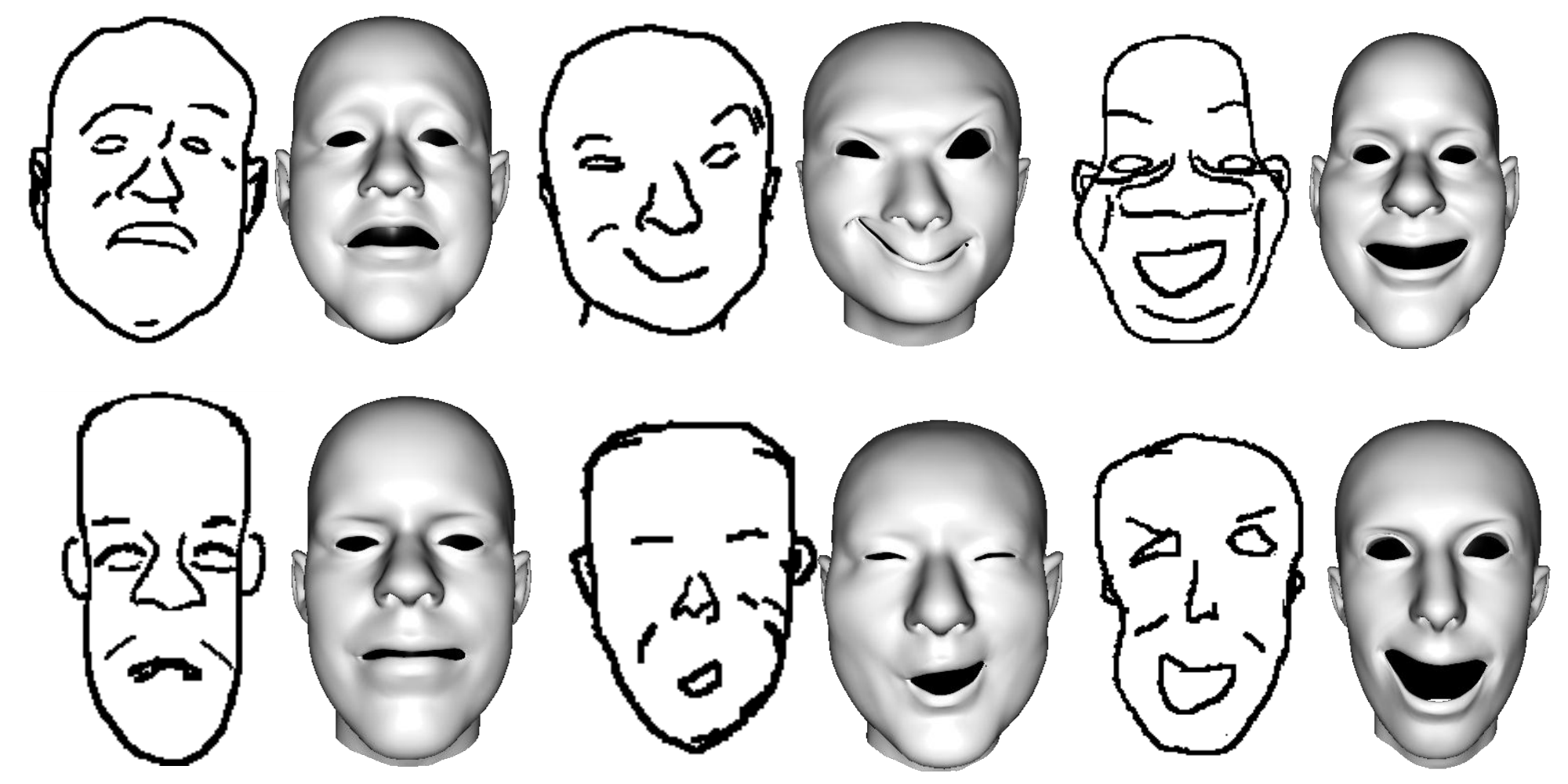}
  \caption{Input sketches are shown side by side with the face models generated by our deep regression network. The first two examples in the first row come from our testing data, which is rendered from 3D models, and all others are freehand drawings. }\label{fig:inferenceresult}
\end{figure}
In the training stage, all the weights in our entire deep network are randomly initialized. To make the training process converge stably, we divide it into three steps: classifier training, $u-v$ regression, and final tuning. We first perform multi-task training for a randomly initialized classification network, which uses softmax layers as its output layers in the two branches. The task is to classify an input sketch into its corresponding identity and expression categories. Therefore, the softmax layer for the identity branch has 600 neurons and that for the expression branch has 25 neurons. In the second step, we fix the weights in the convolutional layers and perform $u-v$ regression by making the two branches predict $u$ and $v$ vectors, respectively. This requires replacing the softmax output layers with $u$ and $v$ regression layers. The number of neurons in these regression layers are set to the number of $u$ and $v$ coefficients we wish to predict. The summed $L_2$ errors of the $u$ and $v$ vectors is used as the loss function during $u-v$ regression. In the final tuning step, we add the vertex loss layer on top of the $u-v$ regression layers, and train the entire network (including all convolutional layers) until convergence.

In addition to the 3D faces in the expanded database, we generate 10,000 extra models by randomly interpolating the $u-v$ parameters of randomly chosen face pairs. These extra models are also rendered into sketches. Such extra data and the original data in the expanded database together form our final data for training and testing. 10\% randomly chosen face models in our final data and their corresponding sketches are used as our testing data.

To perform data augmentation on the rendered sketches, which serve as a large portion of the input images during network training, we first add random noise to the viewing parameters, and then perform random line removal and deformation as in \cite{yu2016sketch} during rendering.

\section{Implementation}


\subsection{Face Modeling from Sketches}
\label{sec:facemodeling}
In the initial sketching mode, our system only updates the 3D face model using our deep regression network and does not perform handle-based shape deformation. The deep network generates the $u$ and $v$ vectors for the bilinear face representation, which further reconstructs all vertex positions for the 3D face mesh. The goal here is creating an initial 3D face that approximately matches the silhouette and feature lines in the freehand drawing sketched by the user. Figure~\ref{fig:inferenceresult} shows sample sketches and their corresponding 3D models inferred by our deep regression network.

In the follow-up sketching mode, the user can improve the 3D face created in the initial sketching mode and make it match user-sketched silhouette and feature lines more precisely. To achieve this goal, our system relies on our deep regression network as well as an existing handle-based shape deformation technique based on the Laplacian~\cite{sorkine2004laplacian}. To make our setup compatible with handle-based shape deformation, as mentioned earlier, we predefine 3D curve handles on a template face mesh representing its silhouette in the frontal view and the contours of facial features. Each curve handle is a vertex sequence on the template mesh. As shown in Figure~\ref{fig:lines} (b), these 3D curves correspond to the silhouette and feature lines in 2D face sketches.
Since all the face meshes in our database share the same topology, the predefined curve handles on the template mesh can be transferred to any face mesh reconstructed from the bilinear face representation.
\begin{figure}[h]
  \includegraphics[width=0.95\columnwidth]{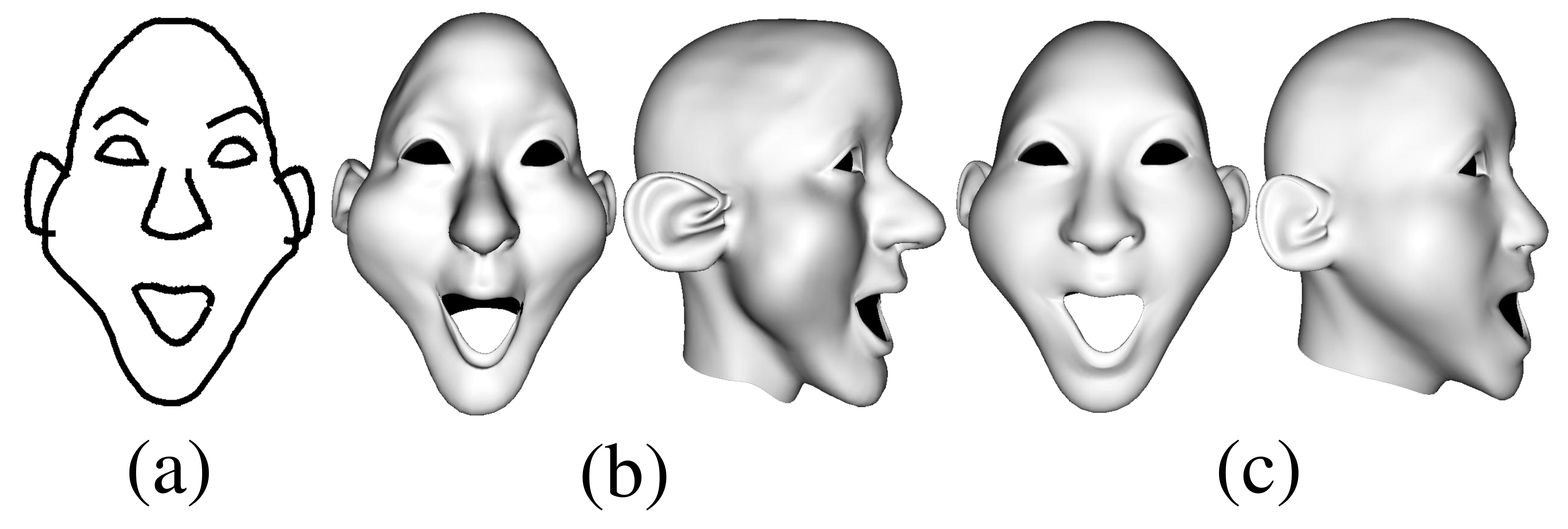}
  \caption{(a) Input sketch, (b) our deformation result with automatic model inference, (c) Laplacian deformation applied to the template model in Figure~\ref{fig:lines}(b).}\label{fig:deformcomp}
\end{figure}

At the beginning of a follow-up sketching session, an initial 2D sketch is automatically drawn by projecting the 3D curve handles on the face model created in the initial sketching mode. As each erasing operation performed on silhouette and feature lines must be followed by a drawing operation, point-wise correspondences between the erased line and the newly drawn line can be defined naturally. After each drawing operation, our system still updates the 3D face model using our deep regression network. In addition, it performs a handle-based deformation so that projected curve handles on the latest face model closely match the silhouette and feature lines in the latest sketch. In this handle-based deformation, the target x-y coordinates of the handle vertices are defined by their corresponding points in the latest sketch while existing z-coordinates of the handle vertices are preserved. In another word, handle-based deformation in the follow-up sketching mode does not change the depth information, which is exclusively updated by our deep regression network. In Figure~\ref{fig:deformcomp}, we show the differences between results obtained with and without performing model inference using our deep network.

The reason we render contours by projecting pre-defined vertex sequences on the 3D face model is that true rendered contours are sensitive to minor geometric variations.
A continuous contour could be broken up into multiple segments on a slightly different model, and relatively large gaps might also occur in-between segments. Such sensitivity negatively affects the performance of 3D model inference.

Handle-based Laplacian deformation is achieved by solving a linear system $A^{T}Ax=A^{T}b$. As curve handles are pre-defined, matrix $A$ is fixed and vector $b$ is updated after each editing operation. Thus, we can perform matrix decomposition on $A^{T}A$ in a pre-processing step, and a solution to the linear system with an updated $b$ can be obtained very efficiently using the precomputed decomposition. These strategies ensure real-time performance of our system.


\subsection{Gesture-Based 3D Face Refinement}
To ensure fluency of user interaction, the key issue in gesture-based 3D face refinement is achieving high accuracy in gesture classification.
\paragraph{Gesture Classification} Unlike existing solutions, convolutional neural networks are chosen for gesture classification. A total of 10 types of gestures have been designed. We first collected the training data from five users, each of whom was asked to supply 200 samples for every type of gestures. The layer composition of our convolutional neural network is organized as follows: 11x11x32 convolution, ReLU, 3x3 max pooling, 5x5x64 convolution, ReLU, 3x3 max pooling, 3x3x16 convolution, ReLU, 3x3 max pooling, 256-d fully connected, 9-d fully connected. These layers are followed by a softmax layer for final classification. The input to the network is a 256x256 binary gesture image. We randomly separated the collected data (10000 images) into 9000 training images and 1000 testing images. Our network achieves 96\% accuracy on the testing images.

\paragraph{Local Deformation} Once handle vertices and an ROI have been selected, handle-based Laplacian deformation~\cite{sorkine2004laplacian} can be performed. In our implementation, we only apply deformation to the surface region enclosed by the ROI. 

\begin{figure*}[t]
  \includegraphics[width=0.98\textwidth]{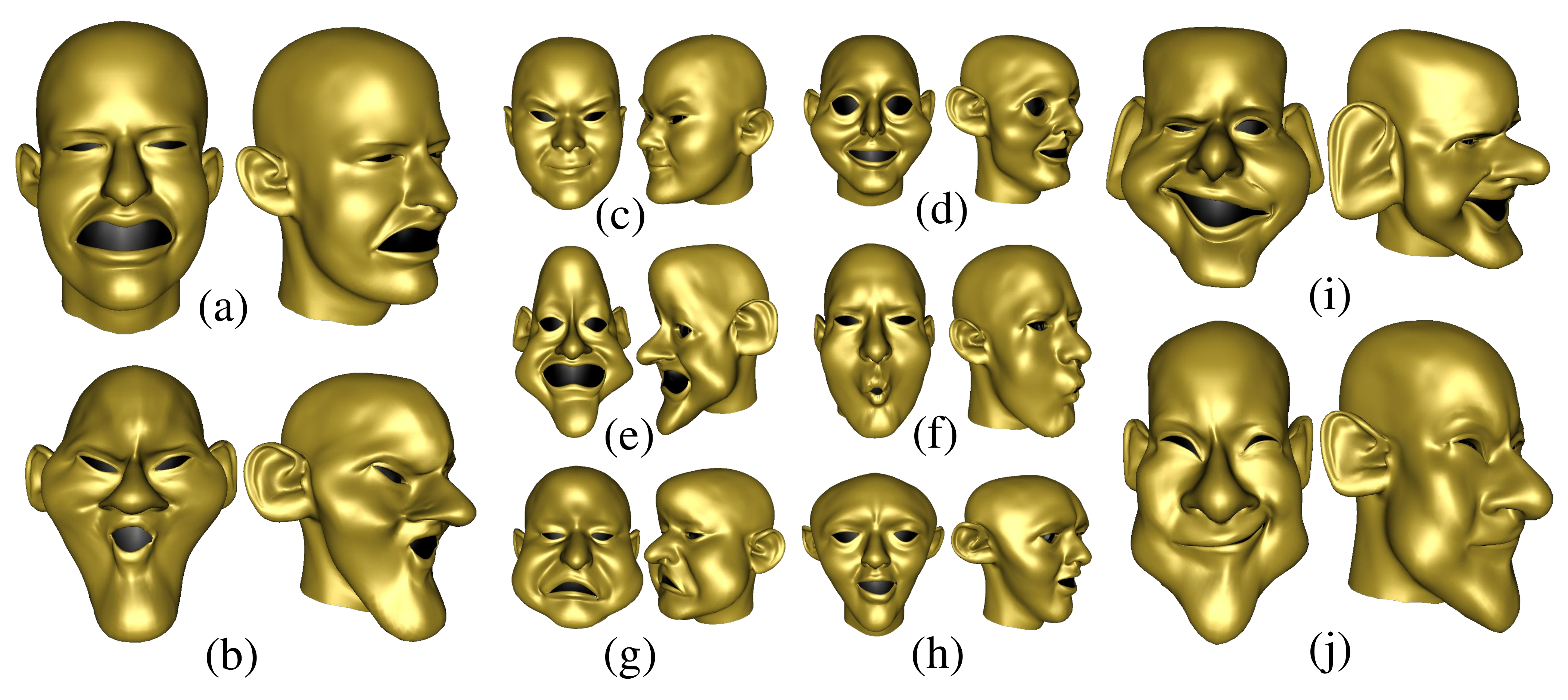}
  \caption{A gallery of results created using our sketching system. On average, each model was created in around 8 minutes.}\label{fig:gallery}
\end{figure*}

\section{Results and User Study}

Our face sketching system has been fully implemented on a PC supporting both mouse and tablet input. We rely on the Caffe deep learning library~\cite{jia2014caffe} to train and test our deep regression network for 3D face model inference on GPUs. As the training process of our deep network consists of three steps, i.e. classifier training, u-v regression and final tuning, the number of iterations in these three steps are set to 500,000, 800,000 and 500,000, respectively. The learning rate is set to 0.001 for the classifier training step and 0.00001 for the other steps. The mini-batch size, momentum parameter and weight decay are set to 50, 0.9 and 0.0005, respectively, for all steps. In both initial and follow-up sketching modes, each face model inference based on our deep regression network takes 50 ms on average on a 3.4GHz Intel processor with a GeForce Titan X GPU. And each handle-based deformation operation takes 41 ms on average.

\begin{figure}[h]
  \includegraphics[width=0.95\columnwidth]{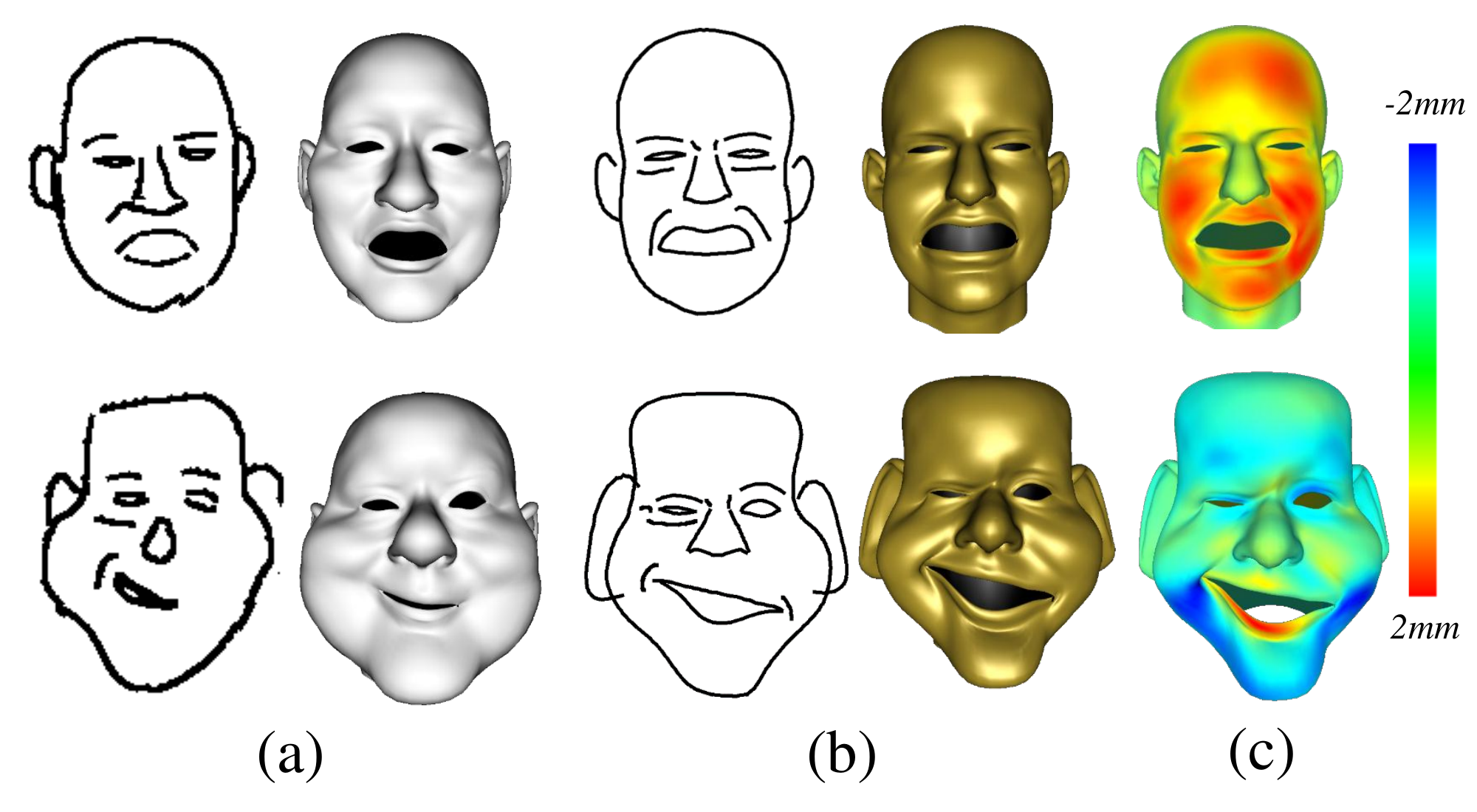}
  \caption{Intermediate results during the creation of model (a) (top) and model (i) (bottom) in Figure~\ref{fig:gallery}. (a) Sketch and model after initial sketching, (b) sketch and model after follow-up sketching, (c) difference map between models obtained with and without wrinkle lines.}\label{fig:intermediate}
\end{figure}

Figure~\ref{fig:gallery} shows a gallery of 3D face models with varying shape and expression. They were created using our system. The interactive process took 8 minutes on average for each model. This includes around 2 minutes on average in the gesture-based refinement mode. Intermediate steps for the creation of two models ((a) and (i) in Figure~\ref{fig:gallery}) are reported in Figure~\ref{fig:intermediate} and Table~\ref{tab:timing}. In Figure~\ref{fig:intermediate}, we show for each example the sketch and model after initial sketching as well as those after follow-up sketching. The timings for every step are reported in Table~\ref{tab:timing}. In addition, we show for each example the difference map between two models generated without and with wrinkle lines (denoted as $M_1$ and $M_2$ respectively) in the follow-up sketching mode. The difference between a vertex pair $p \in M_1$ and $q \in M_2$ is measured by a signed distance, where the value is $||p-q||$ and the sign is defined by $sign(\vec{pq} \cdot \vec{n})$ ($\vec{n}$ is the normal at vertex $p$). These difference maps shows how wrinkle lines influence vertex displacements. More results from first-time users can be seen in Section~\ref{sec:userstudy}.

\begin{table}[h]
\normalsize
\centering
\caption{Timings for intermediate steps during the creation of models (a) and (i) in Figure~\ref{fig:gallery}.}
\label{tab:timing}
\begin{tabular}{|l|c|c|l|}
\hline
                   & \multicolumn{1}{l|}{\textit{Initial Sketching}} & \multicolumn{1}{l|}{\textit{Follow-up Sketching}} & \textit{Refinement} \\ \hline
\textit{model (a)} & 30 seconds                                      & 3 minutes                                         & 1 minute            \\ \hline
\textit{model (i)} & 40 seconds                                      & 7 minutes                                         & 3 minutes           \\ \hline
\end{tabular}
\end{table}

\begin{figure}[h]
  \includegraphics[width=0.95\columnwidth]{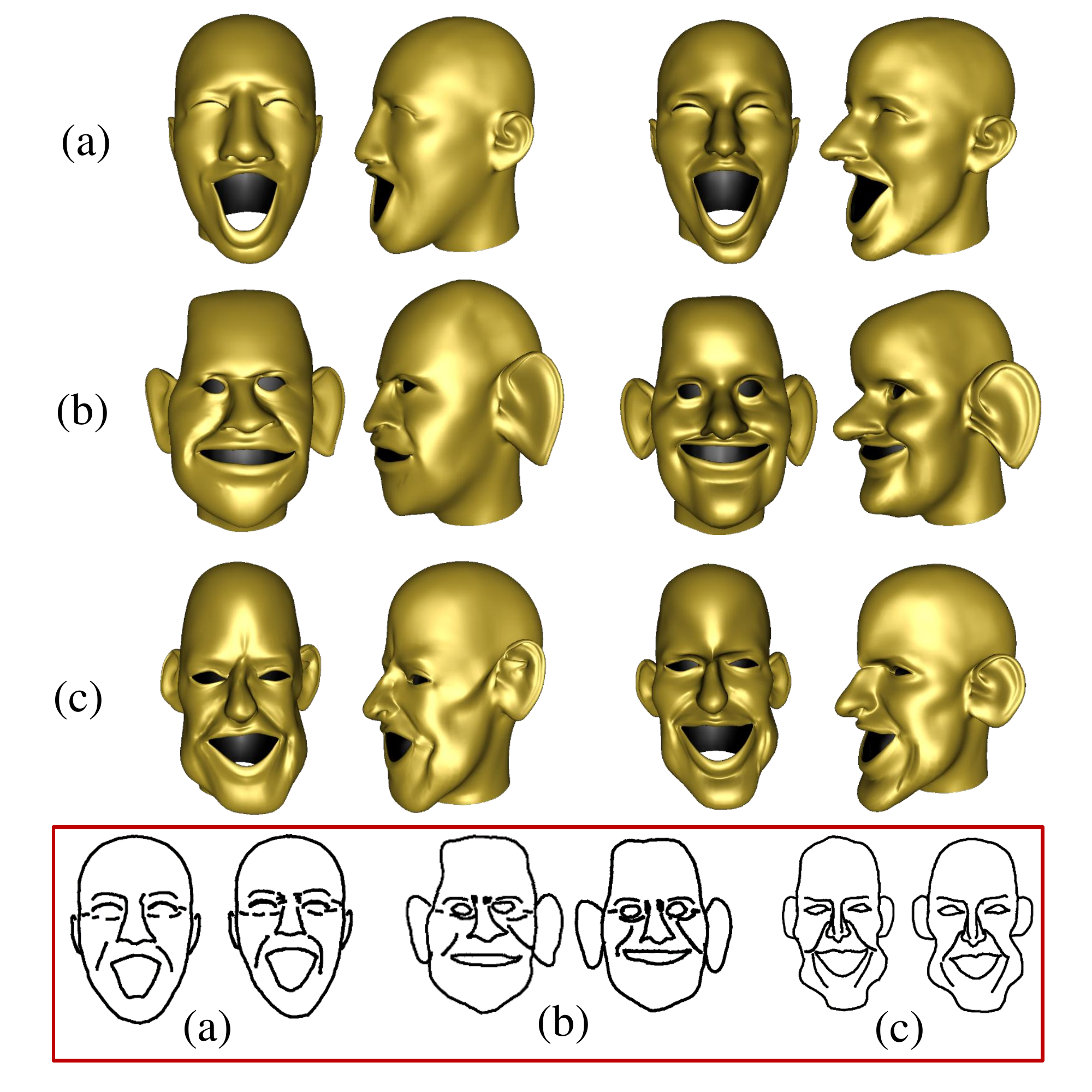}
  \caption{The first three rows show three pairs of models created by users with two different systems in the first stage of our user study. Each row shows the model generated using the deformation-only system and the one using our system. The last row with a red box shows final sketches in the follow-up sketching mode. Within each group, the left and right sketches were drawn in the deformation-only system and our system, respectively. }\label{fig:study1}
\end{figure}

\subsection{User Studies on the Interface}
\label{sec:userstudy}
\paragraph{Stage I: User Experience} We invited 12 amateur users, 8 men and 4 women, to evaluate our system. All participants are graduate students aged 22 to 29. 9 of them have very limited or no experiences about 3D modeling and 2 of them know a little. Only one of them is familiar with 3D modeling. Two of them have more than five years of drawing experiences and all others reported very limited knowledge about drawing. A 15-minute tutorial on how to use our system was given to every participant prior to the modeling session.

Each participant was given a 2D portrait or caricature face as reference, and asked to create a 3D model with a similar shape and expression using our system. Note that the created 3D model was not required to strictly follow the reference image and differences were allowed. We also created another deformation-only user interface supporting face modeling using handle-based interactive mesh deformation only~\cite{sorkine2004laplacian,nealen2007fibermesh,nealen2007sketch}. The follow-up sketching mode and gesture-based refinement mode are still available in this deformation-only interface. The user can perform the same set of interactive operations in these two modes as in the corresponding modes of our proposed interface. While Laplacian deformation using predefined handles is still supported, the main difference is that deep networks are not used for 3D model inference. To validate the effectiveness of deep learning based model inference, each participant was asked to repeat the same task twice independently using our system and the deformation-only interface. The initial sketching mode of our interface was skipped for fair comparison. In the follow-up sketching mode of our system, all participants start modeling from an initial sketch, which was rendered from a template face model. During this modeling session, participants can choose to perform gesture-based 3D model refinement after sketching. This modeling session terminates after 15 minutes or the participant becomes satisfied with the face model displayed on the screen, whichever is earlier.

\begin{figure}[h]
  \includegraphics[width=0.95\columnwidth]{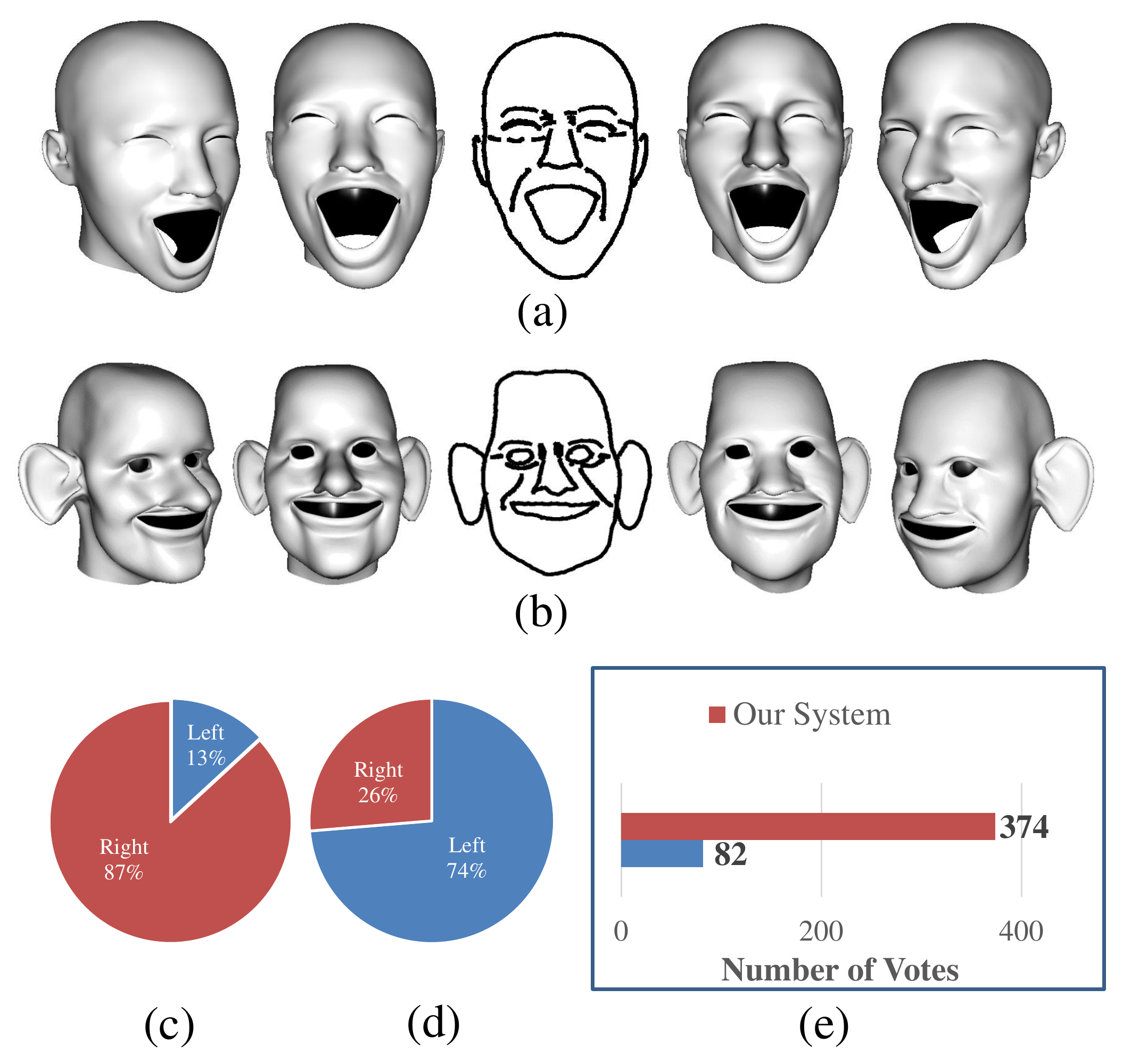}
  \caption{(a) and (b) are two sample questions in the evaluation stage of our user study that correspond to (a) and (b) in Figure ~\ref{fig:study1} respectively. (c) and (d) are voting results corresponding to (a) and (b), respectively. The right model in (a) and the left model in (b) were generated from our system. (e) shows the overall voting result from 38 participants over 12 questions.}\label{fig:study2}
\end{figure}

We asked participants of the above user experience study to complete a short questionnaire that has two questions: 1) Which interface helps users create better face models? 2) Are the wrinkle lines helpful in producing better results? In addition to these questions, we also recorded the amount of time every participant spent in every possible interaction mode of each interface. In summary, all participants have agreed that our system enables users to generate better results. 9 of them agree that wrinkle lines are very helpful and 3 of them feel they are helpful to a certain extent. When using the deformation-only interface, none of the participants managed to finish early, and around 80\% of the time was spent on model refinement. In contrast, a participant on average only spent 10 minutes to complete the task within our system, and around 10\% of the time was spent in the refinement mode. Figure~\ref{fig:study1} shows a subset of the results, and all the remaining results are given in the supplemental materials.

\begin{figure}[h]
  \includegraphics[width=0.95\columnwidth]{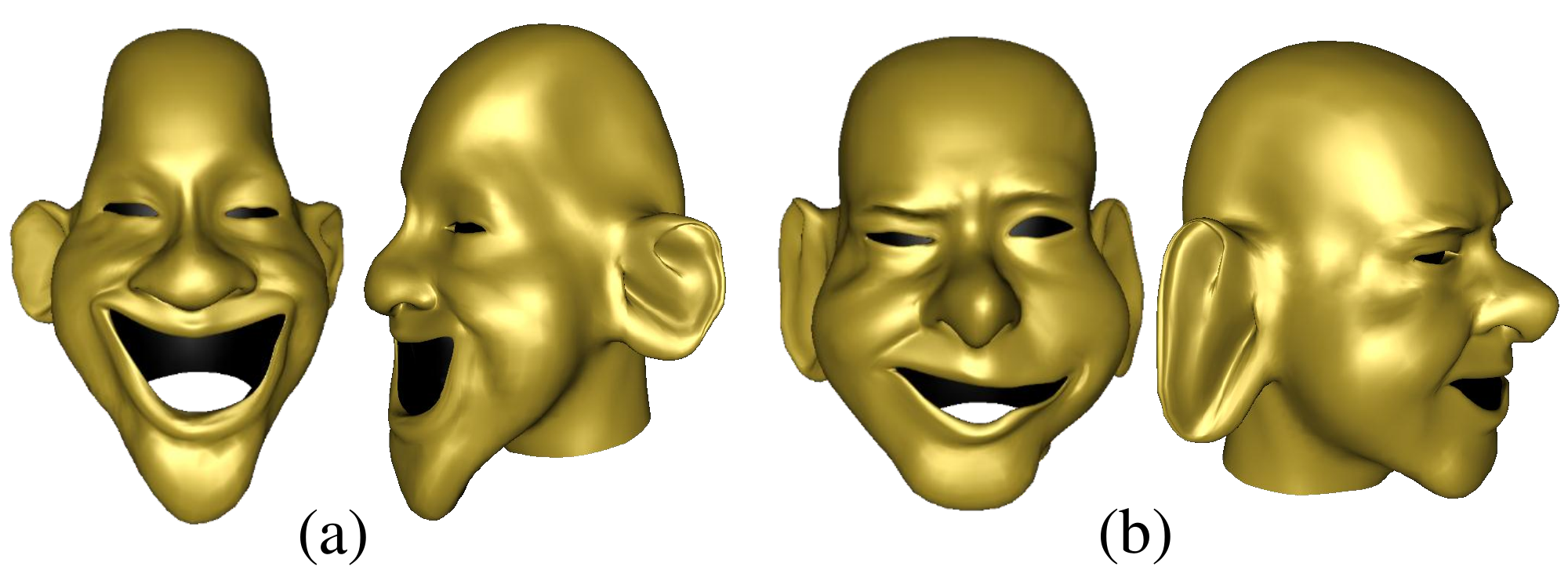}
  \caption{The right model in Figure~\ref{fig:teaser} and model (i) in Figure~\ref{fig:gallery} were used as reference models by a skilled artist. Shown here are corresponding results created by the same artist in 10 minutes using ZBrush. }\label{fig:zbrush}
\end{figure}

\paragraph{Stage II: Evaluation} To further verify that deep learning based 3D model inference can help create better results, we carried out a second stage of the user study. In this stage, we invited 38 additional subjects, who had not participated in the modeling stage, to compare the results created using two different interfaces in Stage I. Every participant needs to look at corresponding models (shown in random order) created in the two interfaces and their associated sketch drawn in Stage I, and was asked to choose the model that looks more natural and better resembles the sketch. The final result is reported in Figure ~\ref{fig:study2}, where two sample questions from the user study are also shown. As reported, among all 456 votes (38 users x 12 questions), our system received 374 votes. All the remaining questions are shown in the supplemental materials.

Though participants in Stage I had been asked to sketch lines in the same way when using the two interfaces, corresponding models still have minor differences in silhouettes and feature lines caused by differences in the sketched lines. To focus our evaluation on the differences in modeling capability, we removed differences caused by freehand sketching by aligning the 3D silhouettes and feature lines on every pair of corresponding models. This was achieved by performing deformation using curve handles defined in Section~\ref{sec:facemodeling}.

\subsection{Comparison with ZBrush} ZBrush is a general-purpose commercial software for creating detailed 3D models. It is a powerful modeling software with a steep learning curve. An informal comparison between our system and ZBrush was conducted on face modeling. We recruited a skilled artist, who had two years of modeling experiences using Zbrush and eight years of drawing experiences. He was asked to use ZBrush to create a 3D model in 10 minutes from each reference model he received. The model on the right in Figure~\ref{fig:teaser} and model (i) in Figure~\ref{fig:gallery} were used as references. Both of them were created within 10 minutes by a user familiar with our system but without any prior drawing experiences. Figure~\ref{fig:zbrush} shows the two models created in ZBrush.

\subsection{Comparisons on 3D Model Inference}
\label{sec:compinference}
Our deep regression network was designed for inferring the coefficients for a bilinear face representation, which in turn reconstructs 3D coordinates of a face mesh. However, there exist many choices on how to infer these bilinear coefficients. We therefore perform a comparison among multiple choices to validate our design.

\begin{figure}
  \includegraphics[width=0.95\columnwidth]{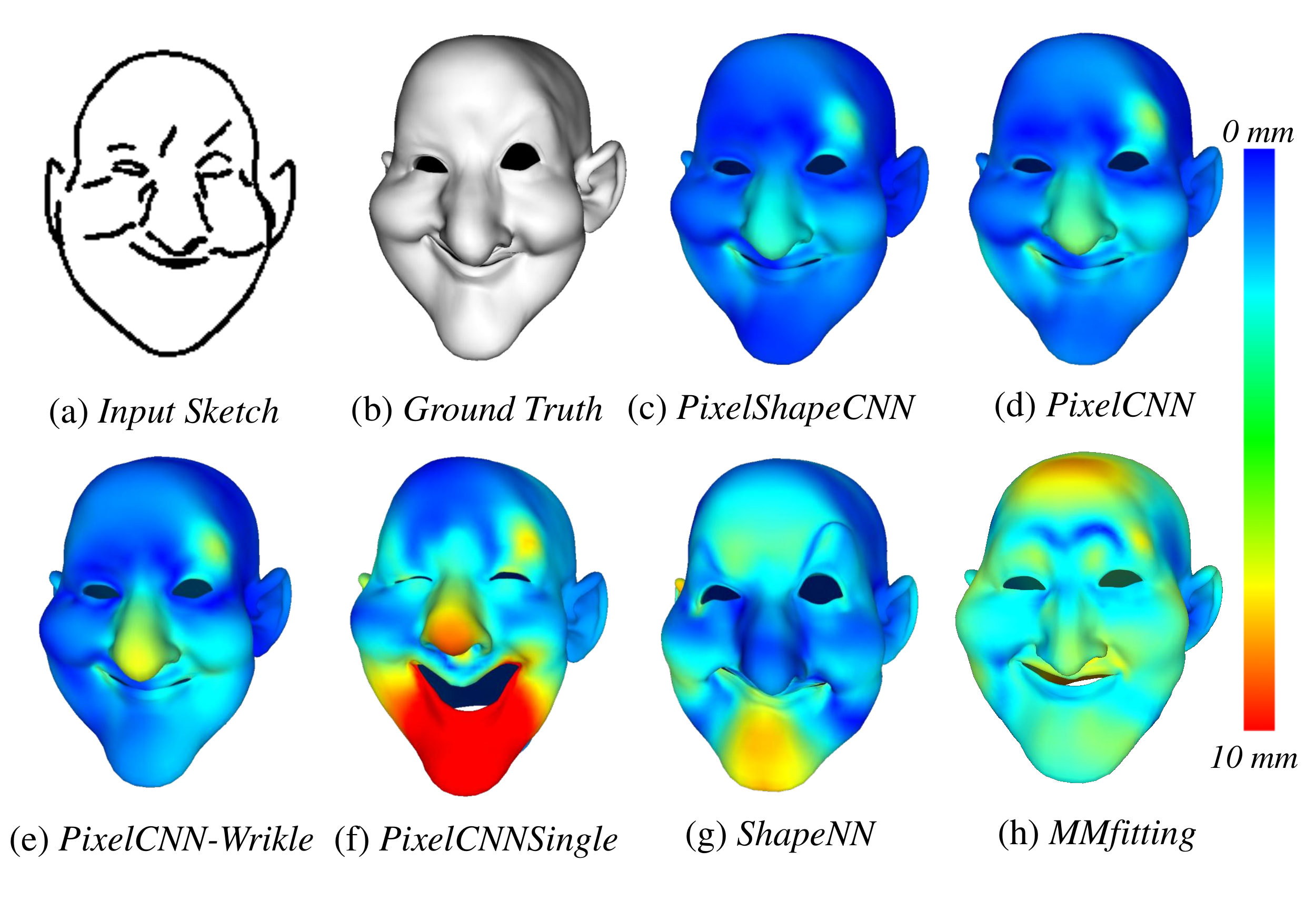}
  \caption{Comparison of error maps among variants of our deep network or algorithm.}\label{fig:comp-model}
\end{figure}

\paragraph{Ablation Study on Network Architecture} We first compare the performance of model inference between our network architecture and its variants. As illustrated in Figure ~\ref{fig:network}, there are three straightforward variants: our final network based on both convolutional layers and 2D bilinear shape encoding (denoted as \emph{PixelShapeCNN}), the network using features from convolutional layers only (denoted as \emph{PixelCNN}), and a regression network only taking 2D bilinear shape encoding as input (denoted as \emph{ShapeNN}). To check the effect of wrinkle lines on model inference, we train another model which has the same network structure as \emph{PixelCNN} but takes pixel-level sketches without wrinkle lines as training images. This model is denoted as \emph{PixelCNN-Wrinkle}. In addition, we also report the performance of a simplified \emph{PixelCNN} which has a single stack of 3 fully connected layers to infer both $u$ and $v$ vectors. This network is denoted as \emph{PixelCNNSingle}. Specifically, the output layers for $u$ and $v$ vectors in this network are simultaneously connected to the last fully connected layer in the stack.

These five networks have been applied to our testing set independently. Given two 3D models with exactly the same vertex connectivity, we use the average distance between their corresponding vertices to define the distance between these two models. Then, the average distance between inferred models and their groundtruth models is reported as the mean error in Table~\ref{tab:comp-error}, where we assume an average face is 200mm tall and the inference errors have been scaled accordingly. It can be verified that our \emph{PixelShapeCNN} achieves the highest accuracy. The mean error attained by \emph{PixelCNNSingle} is significantly higher probably because the regression functions for the identity and expression modes are highly incompatible and sharing the same set of fully connected layers give rise to severe interference between them. Meanwhile, using convolutional features only without 2D bilinear shape encoding increases the mean error by 8.8\%, and skipping wrinkle lines in the input sketches increases the mean error by 28.9\%. On the other hand, using 2D bilinear shape encoding only without convolutional features increases the mean error by 64.7\%, which indicates the importance of convolutional features.
We show an example of the face models obtained by all these variants in Figure~\ref{fig:comp-model}, where color maps are used to visualize the error distribution across all vertices.

\begin{table}[t]
\begin{center}
\begin{tabular}{@{}lccc@{}}
\hline
network                          & mean error (mm)  \\ \hline
\textit{PixelShapeCNN}           & \textbf{2.04}    \\
\textit{PixelCNN}                & 2.22     \\
\textit{PixelCNN-Wrinkle}        & 2.63     \\
\textit{ShapeNN}                 & 3.36     \\
\textit{PixelCNNSingle}          & 7.83     \\
\textit{MMfitting}               & 6.06     \\
\hline
\end{tabular}
\end{center}
\caption{Ablation study on network architecture.}
\label{tab:comp-error}
\end{table}

\paragraph{Morphable Model Fitting} As the silhouette and feature lines of an input sketch correspond to a set of key vertices on the 3D mesh, a direct solution for $u-v$ inference performs a morphable model fitting as in \cite{cao2014facewarehouse} to minimize the errors between projections of the key vertices and their corresponding 2D positions in the sketch. The result is denoted as \emph{MMfitting}. In our context, we assume an orthographic projection, and thus the camera projection matrix is known. The mean error across all testing sketches is reported in Table~\ref{tab:comp-error}, and Figure~\ref{fig:comp-model} also illustrates the error map of an example obtained this way. Morphable model fitting finds the optimal model by fitting points on the 2D sketch. It does not use any depth constraints since such constraints are unavailable. As a result, the accuracy of its depth estimation is much lower than that of fitting 2D points. As shown in Table 2, the error introduced by MMfitting is three times larger than the error of our method. As shown in Figure 15(h), the visual quality of MMfitting is also less than satisfactory (e.g. erroneous wrinkles on the forehead).

\begin{figure}[h]
  \includegraphics[width=0.95\columnwidth]{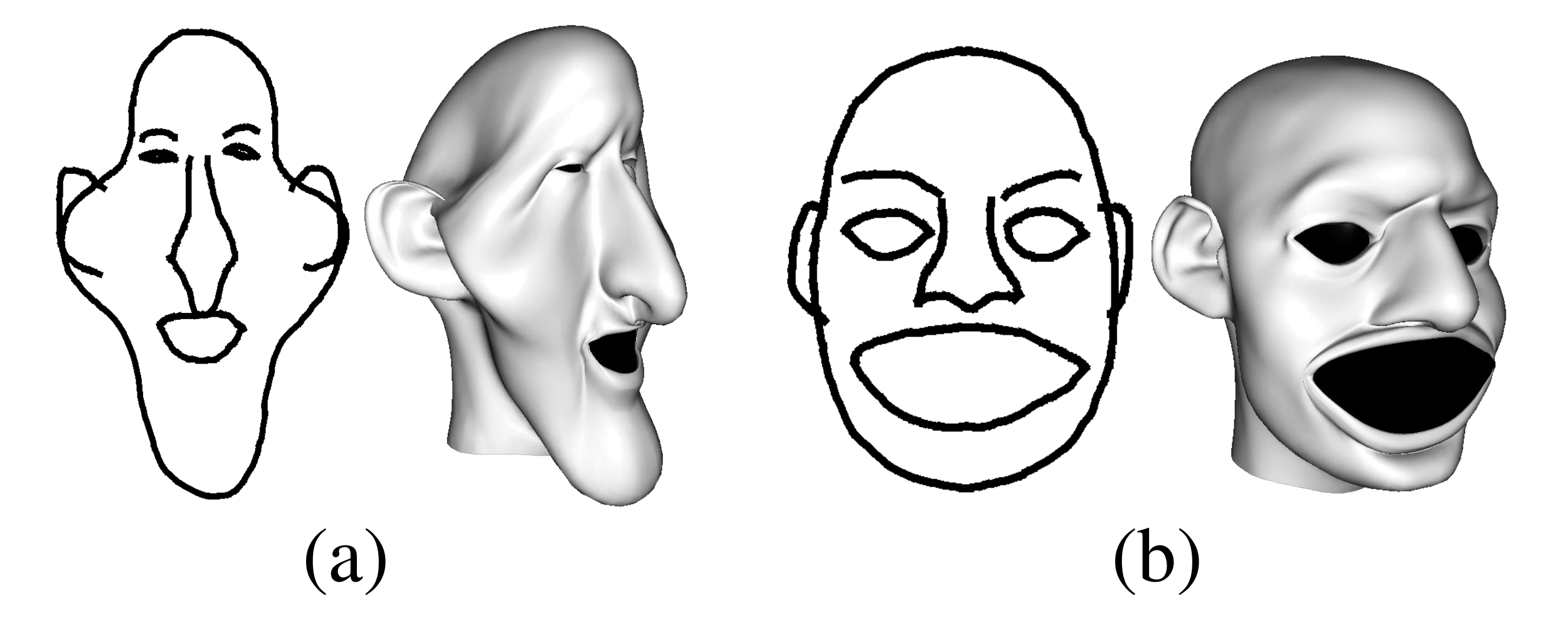}
  \caption{Limitations of our database and system. Our system generates unnatural results when given inconsistent exaggeration of face parts. }\label{fig:limitation}
\end{figure}

\section{Conclusions and Discussion}
In this paper, we have presented a deep learning based sketching system for 3D face and caricature modeling. The sketching interface in this system allows the user to draw freehand imprecise yet expressive 2D lines representing the contours of facial features. A novel CNN based deep regression network is designed for inferring 3D face models from 2D sketches. Our network has two independent branches of fully connected layers generating independent subsets of coefficients for a bilinear face representation. Our system also supports gesture based interactions for users to further manipulate initial face models. Both user studies and numerical results indicate that our sketching system can help users create face models quickly and effectively. A significantly expanded face database with diverse identities, expressions and levels of exaggeration has also been constructed to promote further research and evaluation of face modeling techniques.

\paragraph{Limitations and future work} As shape exaggeration was performed on all face parts consistently when we built our database, our system creates unnatural results when the level of exaggeration in shape and expression is highly inconsistent across different parts. Two such examples are shown in Figure ~\ref{fig:limitation}. Alleviating this problem by further expanding our database is one of the directions of our future work. Moreover, as morphable models are not able to create geometric details such as wrinkles at novel locations, inferring such details from sketches is another future direction. To reach this goal, it is important to move beyond morphable models and use a deep network, such U-Nets~\cite{ronneberger2015u} and GAN~\cite{goodfellow2014generative}, well suited for pixel-to-pixel prediction.

\begin{acks}

The authors would like to thank the reviewers for their constructive comments, and the participants of our user study for their precious time.

\end{acks}

\citestyle{acmauthoryear}
\setcitestyle{square}
\bibliographystyle{ACM-Reference-Format}
\bibliography{ske2face}

\end{document}